\documentclass[showpacs,aps,prl,twocolumn,superscriptaddress,longbibliography]{revtex4-1} 
\usepackage{dcolumn}
\usepackage{bm}
\usepackage[colorlinks,hyperindex, urlcolor=blue, linkcolor=blue,citecolor=black, linkbordercolor={.7 .8 .8}]{hyperref}
\usepackage{graphicx}
\usepackage{tabularx}
\usepackage{amsfonts}
\usepackage{amsmath}
\usepackage{amssymb}
\usepackage{amsbsy}
\usepackage{setspace}
\usepackage{nicefrac}
\usepackage{xfrac}
\usepackage{bm}

\newcommand{\vf}{$v^{ee}_F$}

\begin{document}

\newcommand{\refs}{/Users/erikku/Documents/Papers3/papers.bib}

\title{Many-particle effects in the cyclotron resonance of encapsulated monolayer graphene}

\author{B.~Jordan Russell}
\affiliation{Department of Physics, Washington University in St.~Louis, 1 Brookings Dr., St.~Louis MO 63130, USA}
\author{Boyi Zhou}
\affiliation{Department of Physics, Washington University in St.~Louis, 1 Brookings Dr., St.~Louis MO 63130, USA}
\author{T.~Taniguchi}
\affiliation{National Institute for Materials Science, 1-2-1 Sengen, Tsukuba, Ibaraki 305-0044, Japan}
\author{K.~Watanabe}
\affiliation{National Institute for Materials Science, 1-2-1 Sengen, Tsukuba, Ibaraki 305-0044, Japan}
\author{Erik A.~Henriksen}
\affiliation{Department of Physics, Washington University in St.~Louis, 1 Brookings Dr., St.~Louis MO 63130, USA}
\affiliation{Institute for Materials Science \& Engineering, Washington University in St.~Louis, 1 Brookings Dr., St.~Louis MO 63130, USA}
\email{henriksen@wustl.edu}

\begin{abstract}
We study the infrared cyclotron resonance of high mobility monolayer graphene encapsulated in hexagonal boron nitride, and simultaneously observe several narrow resonance lines due to interband Landau level transitions. By holding the magnetic field strength, $B$, constant while tuning the carrier density, $n$, we find the transition energies show a pronounced non-monotonic dependence on the Landau level filling factor, $\nu\propto n/B$. This  constitutes direct evidence that electron-electron interactions contribute to the Landau level transition energies in graphene, beyond the single-particle picture. Additionally, a splitting occurs in transitions to or from the lowest Landau level, which is interpreted as a Dirac mass arising from coupling of the graphene and boron nitride lattices.
\end{abstract}

\date{\today}

\maketitle

In the presence of a magnetic field, the linear dispersion of electrons in graphene becomes quantized into a set of discrete Landau levels (LLs) with energies exhibiting a highly unusual square-root dependence on both the field strength and the LL index~\cite{CastroNeto:2009wq}. Infrared optical probes of cyclotron resonance (CR) transitions between these LLs can be used to explore this spectrum and indeed both the field and index dependence were found~\cite{Sadowski:2006iz,Jiang:2007hm,Deacon:2007,Orlita:2010tk}. In contrast to more conventional 2D systems having a parabolic dispersion---in which evenly-spaced LLs lead to only a single CR peak---the index dependence in graphene enables many transitions to occur simultaneously, yielding richer optical spectra. Moreover, in these parabolic systems long wavelength light couples only to the center-of-mass motion, so that in principle electron-electron interactions do not impact the CR; in actual practice disorder or non-parabolic band structures can render the CR weakly sensitive to interactions~\cite{Kohn:1961uk,Wilson:1980vh,Englert:1983ta,Kallin:1985ud,Heitman:1986vz,Chou:1988tn,Nicholas:1989wd,Cheng:1990vr,Henriksen:2006ey}. Intriguingly, the linear dispersion of graphene should enable CR that is \emph{intrinsically sensitive} to interactions~\cite{Jiang:2007hm,Bychkov:2008gi,Henriksen:2010ci,Shizuya:2010go,Chen:2014bz,Faugeras:2015it}, leading to widespread anticipation that optical probes in graphene will provide a novel window on the many-body problem~\cite{Iyengar:2007cq,Shizuya:2011kr,Roldan:2010fu,Bisti:2011dq,Lozovik:2012uu,Shovkovy:2016fv,Sokolik:2017cn}. 

Here we report the unambiguous observation of many-particle interaction contributions to the single-particle infrared CR energies in high mobility encapsulated monolayer graphene, manifesting as a non-monotonic variation of the effective Fermi velocity of the charge carriers in a magnetic field. The highest quality graphene devices in which electron-electron interaction effects are expected to be strongest are generally microscopic with typical dimensions of order $10~\mu$m~\cite{Dean:2011ks,young:2012}, and are therefore difficult to study with traditional infrared magneto-spectroscopy techniques. Improving on our prior work~\cite{Jiang:2007hm,Henriksen:2010ci}, we have developed a capability for probing such microscopic devices via simultaneous measurements of optical and electronic properties at high magnetic field, enabling pursuit of the physics of interacting electrons in a new regime.

In a single-particle picture, the energies of interband Landau level transitions in graphene are given by
\begin{equation}
\Delta E_{mn} ~=~ v_F \sqrt{2 e \hbar B} \left( \sqrt{|m|} + \sqrt{|n|} \right), \label{eq1}
\end{equation}
\noindent where $m$ and $n$ index the initial and final levels and $v_F\approx10^6$ m/s is the carrier Fermi velocity which is nominally constant. However electron-electron interactions can contribute an additional energy $\delta E_{ee} = C_{mn} e^2 / l_B$, where $C_{mn}$ may depend on the initial and final states $m$ and $n$ and on the field as well~\cite{Kallin:1985ud,Bychkov:2008gi,Shizuya:2010go}; $l_B = \sqrt{\hbar/eB}$ is the magnetic length with $\hbar$ the reduced Planck's constant and $-e$ the electron charge. Because both $\Delta E_{mn}$ and $e^2 / l_B$ vary with $\sqrt{B}$, by changing the field alone it is difficult to clearly distinguish between the single- and many-particle contributions. Combining these two contributions yields
\begin{equation}
\Delta E_{mn}^{total} ~=~ \Delta E_{mn} + \delta E_{ee} ~=~ \left( v^{ee}_F / v_F \right) \Delta E_{mn}~,\label{eq2}
\end{equation}
\noindent where the ``effective'' Fermi velocity $v^{ee}_F{\geq}v_F$ accounts for the interaction contribution~\cite{Jiang:2007hm,Elias:2011ev,Chae:2012jk,Yu:2013ku,Chen:2014bz,Faugeras:2015it}. Thus at constant field, any variation in $v^{ee}_F$ reflects changes arising from many-particle interactions.

Therefore, in this work we study CR transitions at \emph{constant} $B$ field and tune the carrier density, $n$, through its dependence on the back gate voltage, $V_g$. This amounts to controlling the LL filling factor, $\nu {=} n h / e B$. While the LL spacings and Coulomb energy are thus held fixed, the physics of correlated electrons in partially-filled LLs may yet impact the CR. Indeed we observe a \emph{clear non-monotonic dependence of \vf~on the filling factor}, with the specific behavior changing for each of the six LL transitions measured.

\begin{figure*}[t]
\includegraphics[width=0.9\textwidth]{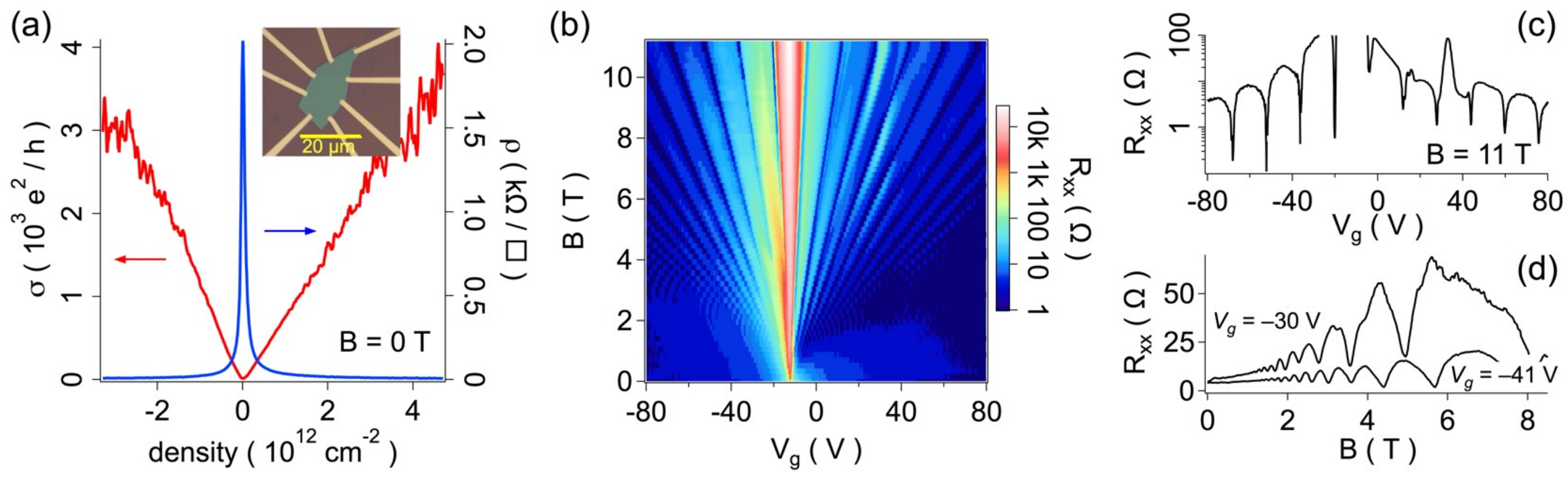}
\caption{Electronic transport at T = 6 K. (a) Zero-field conductivity and resistivity vs carrier density, $n$; the inset shows an optical micrograph of the $250~\mu$m$^2$ encapsulated graphene device used in this work. (b) Magnetoresistance vs gate voltage, $V_g$. Note that up to $V_g = \pm 80$ V, there is no evidence of additional Dirac peaks sometimes observed in transport of hbn-encapsulated graphene~\cite{Moon:2013cp,Ponomarenko:2013hl,Dean:2013bv,Hunt:2013ef}. (c) Magnetoresistance traces at constant field $B=11$ T, and (d) Shubnikov-de Haas oscillations. \label{transport}} 
\end{figure*}

Infrared magneto-spectroscopy was performed at T = 6 K in a cryostat modified to admit light from below the sample~\footnote{See Supplemental Material, which contains additional Refs.~\cite{Mueller:2009ez,Kuzmenko:2005jh,pasm,Vozmediano:2011eg}, for details of the experimental setup, curve fitting, and complementary data and analysis}. Broadband light sourced from a SiC globar in a Fourier-transform infrared spectrometer passes through a KBr window and custom parabolic reflecting optics, en route to detection by a composite Si bolometer affixed to a compound parabolic collector. Electronic transport data were acquired via standard low-frequency lockin techniques during the same experimental run as the infrared measurements. 

The sample consists of graphene encapsulated between two hexagonal boron nitride (hbn) crystals with a total area of $250~\mu$m$^2$, which is placed on a lightly-doped, oxidized Si wafer, and contacted along the edges by Ti/Al wires~\cite{Wang:2013ch} (see Fig.~\ref{transport}a). The five layers of the finished device---Si/oxide/hbn/g/hbn---vary in thickness and dielectric properties. Infrared transmission through this structure results in distortions of the Lorentzian lineshapes, so that the actual central energies and widths of each resonance may differ somewhat from values naively extracted from the transmission lineshapes~\cite{yeh2005optical,Henriksen:2010ci}. To account for these effects, we obtain the resonance energies and linewidths from a nonlinear fitting procedure that accounts for multiple reflections in a thin-film stack~\cite{Note1}.

\begin{figure}[t]
\includegraphics[width=0.9\columnwidth]{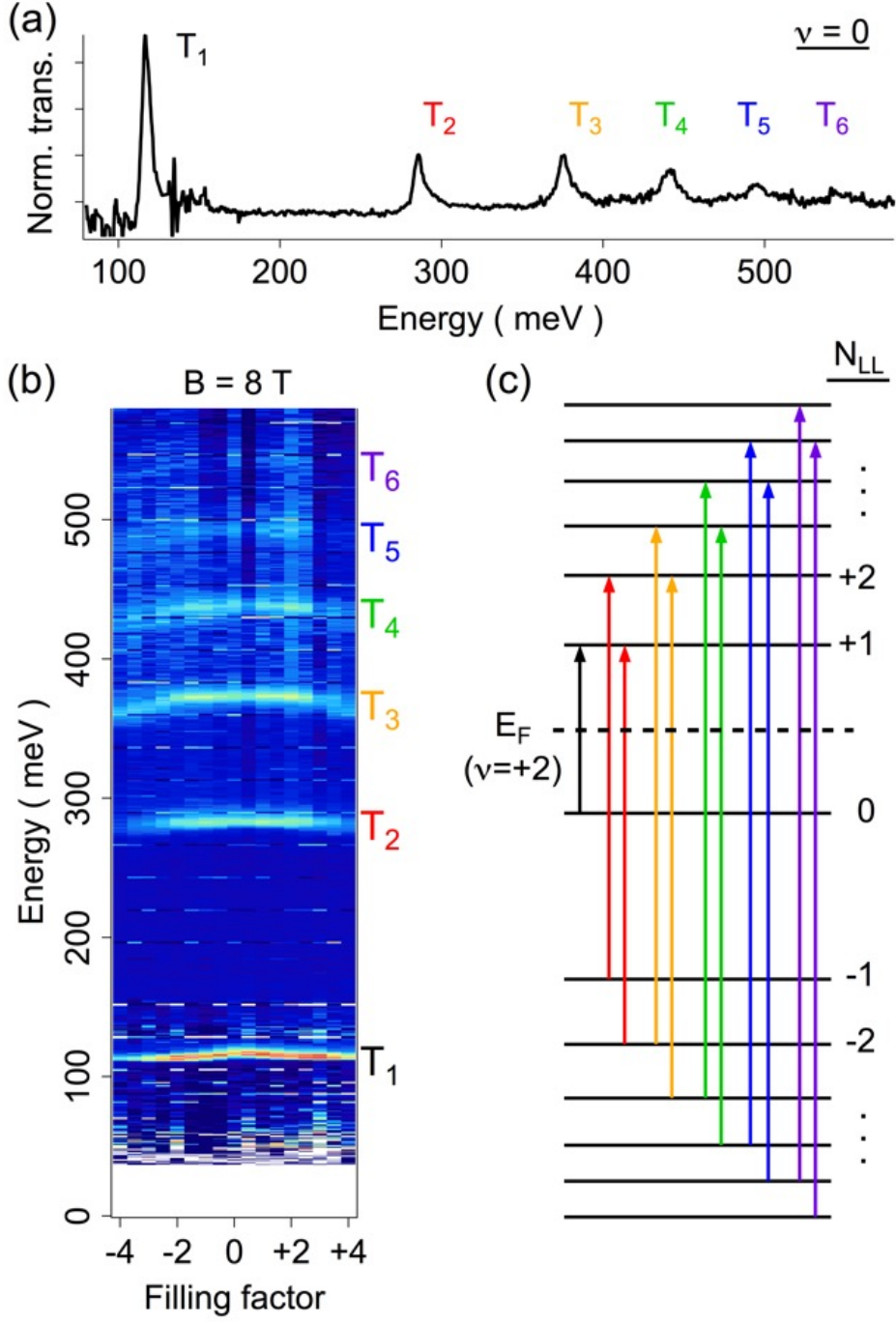}
\caption{Interband Landau level transitions in monolayer graphene at $B{=}8$ T. (a) Typical normalized transmission, at $\nu{=}0$, showing six resonances. The labels identify the same transitions in all plots. (b) Colormap of resonances vs $\nu$. The color scale is expanded above 290 meV to better show lower intensity peaks. (c) Schematic of Landau levels,  and allowed interband transitions (arrows in color) at $\nu{=}{+}2$~\cite{Gusynin:2007cz}.  \label{cr}} 
\end{figure}

The device exhibits hallmarks of high quality graphene as demonstrated by the electronic transport. Figure~\ref{transport}a shows the conductivity and resistivity vs carrier density. Linear fits yield an electron (hole)-side mobility of 190,000 (290,000) cm$^2$/Vs. Although the graphene sheet has two interfaces with hbn crystals, there is no evidence here of the satellite Dirac peaks that might indicate the presence of moir\'e superlattices or Hofstadter butterfly physics~\cite{Moon:2013cp,Ponomarenko:2013hl,Dean:2013bv,Hunt:2013ef}. The magnetoresistance, $R_{xx}$, vs both the magnetic field, $B$, and $V_g$ is shown in Fig.~\ref{transport}(b), with additional traces at constant field or $V_g$ shown in Fig.~\ref{transport} (c) and (d), respectively. The integer quantum Hall minima in $R_{xx}$ are deeper at negative $V_g$, correlating with the higher hole mobility. No symmetry breaking of the four-fold degenerate Landau levels is observed~\cite{young:2012}, likely due to the elevated temperature. The transport was negligibly impacted by exposure to ir light. 

\begin{figure}[b]
\includegraphics[width=0.9\columnwidth]{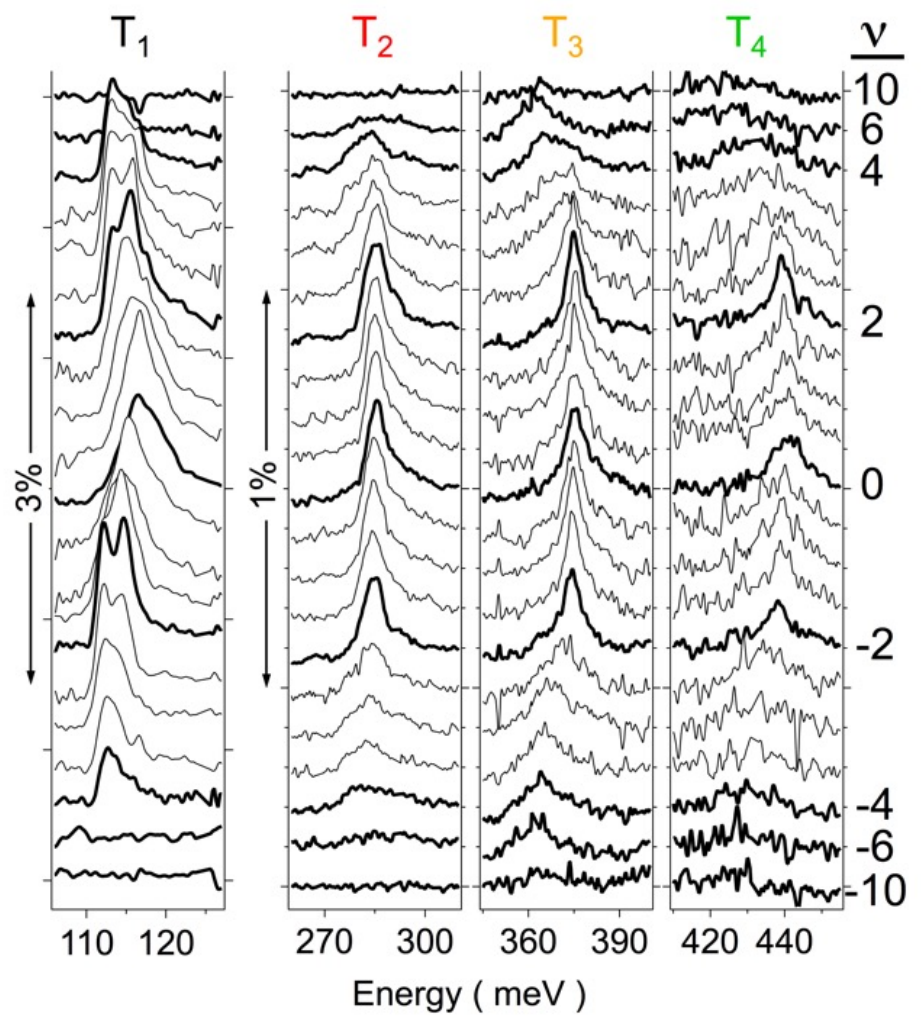}
\caption{Filling factor dependence of the first four interband transitions. For $|\nu|{>}2$---when the Fermi level lies outside the lowest Landau level---\emph{all} transitions are seen to broaden and undergo shifts in the central energy. Nonetheless good fits are readily made~\cite{Note1}. Note the larger vertical scale for T$_1$. \label{trans}} 
\end{figure}

Infrared spectra were acquired at $B{=}5$, 8, and 11 T, as a function of the filling factor. The raw ir traces taken at $|\nu|{\leq}10$ are normalized to the transmission at $\nu{=}22$ and plotted as $1-S(\nu) / S(\nu{=}22)$, where $S$ is the raw transmission signal. Thus at a given $\nu$ value, resonant absorption of ir light leads to reductions in $S$ which appear as peaks in Figs.~\ref{cr} and \ref{trans}. A representative trace taken at $\nu{=}0$ and $B{=}8$ T is shown in Fig.~\ref{cr}a. Several resonances labeled T$_1$ through T$_6$ are clearly visible, spanning the accessible range of mid-infrared frequencies. A series of such traces taken over a range of $\nu$ values at $B{=}8$ T is combined into a colormap in Fig.~\ref{cr}b. With reference to the Landau level schematic in Fig.~\ref{cr}c---in which the allowed interband transitions at $\nu{=}{+}2$ according to the selection rule $n{=}|m|{\pm}1$ are shown~\cite{Gusynin:2007cz}---the resonance energies in Fig.~\ref{cr}b are seen to hew closely to the square root dependence on LL index described by Eq.~\ref{eq1}. 

Figure~\ref{trans} shows the detailed $\nu$-dependence of the four lowest transitions, T$_1$ ($0\to1$ and $-1\to0$), T$_2$ ($-1\to+2$ and $-2\to+1$), T$_3$ ($-2\to+3$ and $-3\to+2$), and T$_4$ ($-3\to+4$ and $-4\to+3$). In Fig.~\ref{trans}, T$_2$, T$_3$, and T$_4$ show similar behavior: for $|\nu|{<}2$, a single peak remains largely unchanged as $\nu$ is varied, but for $|\nu|{>}2$ the peak begins to broaden and show shifts of the central energy; this effect is most pronounced for T$_3$ and T$_4$. Note that while the lower intensity resonances become difficult to see on the same scale as full strength peaks, fits to the data remain robust~\cite{Note1}. Meanwhile T$_1$ exhibits a more elaborate behavior, with a maximum in the resonance energy occurring at $\nu{=}0$, and the appearance of a splitting centered near $\nu{=}{-}2$ and $\nu{=}{+}2.5$. Except for the splitting, the behavior of T$_1$ is consistent with the findings of Ref.~\cite{Henriksen:2010ci} which were interpreted as an interaction-induced gap at half-filling of the $N_{LL}{=}0$ level~\cite{young:2012}.

\begin{figure*}[t]
\includegraphics[width=\textwidth]{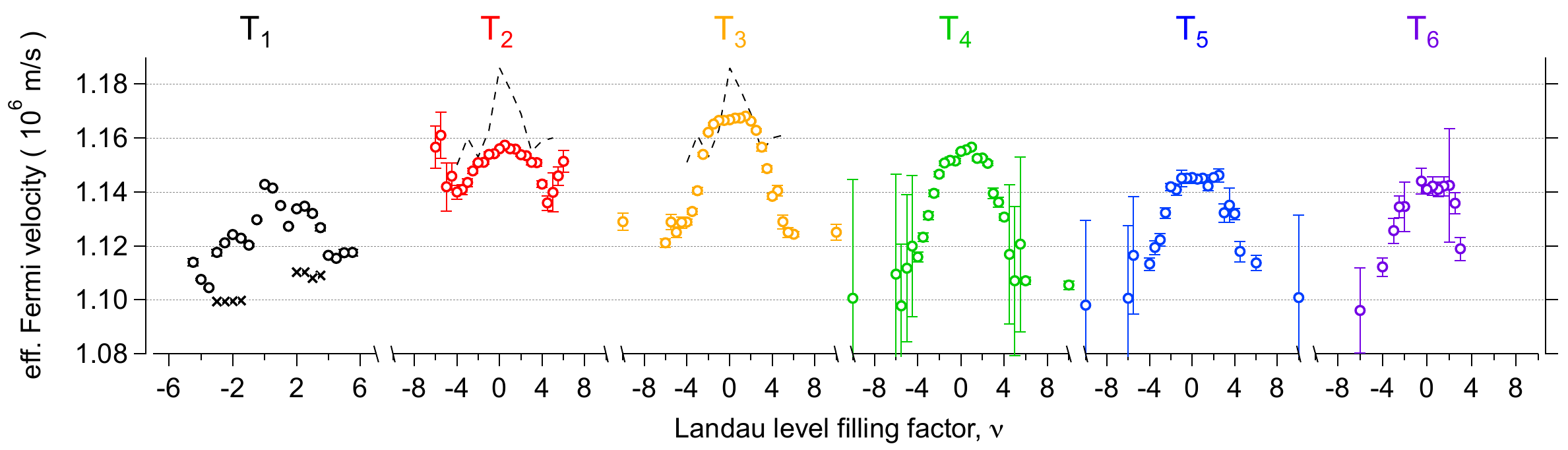}
\caption{Effective Fermi velocity at $B{=}8$ T for all six transitions as a function of filling factor, $\nu{=}n h / e B$. The black dashed lines in T$_2$ and T$_3$ are calculated from the theory of Ref.~\cite{Bychkov:2008gi}. \label{vf}} 
\end{figure*}

Figure~\ref{vf} shows the detailed behavior of the effective Fermi velocity, directly extracted from the CR energies by $v^{ee}_F= \Delta E^{meas}_{mn} / (\sqrt{2 e \hbar B}~(\sqrt{|m|}+\sqrt{|n|}~) )$, as a function of $\nu$~\footnote{This approach differs slightly from other works, where the Fermi velocity is studied in ratio to that of the $0\to1$ transition: $v_F(T_j)/v_F(T_1)$~\cite{Shizuya:2010go,Chen:2014bz}. Since here T$_1$ varies in a distinct fashion from the other transitions, indicating different physics is at work, the meaning of that ratio is not obvious and so we plot each \vf~for each transition independently.}. Several intriguing features appear: (1) \vf~shows a non-monotonic dependence on filling factor; (2) the behavior of \vf~falls in two groups: T$_3$ through T$_6$ show a generally similar response as $\nu$ changes, while T$_1$ and T$_2$ each display a unique dependence on $\nu$; (3) the peak values of \vf~also vary non-monotonically with the \emph{transition number}; and (4) \vf~is also found to decrease monotonically with increasing magnetic field.

We address each of these in turn:

\noindent (1) At constant $B$, the single-particle LL separations and bare Coulomb interaction $e^2/l_B$ are fixed, therefore the variations in \vf~must arise from changes in the many-particle interactions experienced by the excited electron and the hole it leaves behind. The density of states at the Fermi level and therefore the effects of many-particle screening will change with filling factor, so the exciton's energy will change as well. 

\noindent (2) The variation of \vf~with filling factor is different for each transition, but the data in Fig.~\ref{vf} can be divided in two groups: transitions T$_3$ through T$_6$ show similar $\nu$-dependence suggesting a common origin, with (i) a peak or plateau for $|\nu|{<}2$ that is centered on $\nu{=}0$; and (ii) a sharp 4-5\% decrease for $|\nu| > 2$. In contrast, transitions T$_1$ and T$_2$ both show a maximum at $\nu{=}0$ but otherwise show distinct behavior.

The existence of these groupings is likely due to the fact that, for almost all filling factors studied, the Fermi level lies in the $N{=}0$ or $\pm1$ LLs which participate in the T$_1$ and T$_2$ transitions. If any of these three lowest levels shift or split when partially filled, the CR transition energies should reflect this. For instance, the sharp minima in T$_2$ at $|\nu|{=}4$ (half-filling of $N{=}\pm1$) are reminiscent of exchange-enhanced gaps seen at half-filling in high-mobility GaAs 2D systems~\cite{Dial:2007wi}. Similarly, a sharp maximum in T$_1$ is seen for $\nu{=}0$ (half-filling of $N{=}0$)~\cite{Henriksen:2010ci}. On the other hand, the T$_3$--T$_6$ transitions comprise electrons excited over the Fermi sea from completely full to completely empty LLs. For these, the enhanced values of \vf~arise purely from many-body interactions with states at the Fermi energy in a distant, partially filled LL~\cite{Bychkov:2008gi,Shizuya:2010go}.

These many-particle contributions to T$_3$--T$_6$ are fairly constant until $|\nu|{>}2$, after which \vf~abruptly decreases. Similar behavior at 11 T emphasizes this decrease is associated with changes in the LL filling rather than a particular density~\cite{Note1}. At $\nu{=}\pm2$, the Fermi level shifts from $N{=}0$ to $N{=}\pm1$, so the decrease may be due to enhanced screening from the presence of a partially-filled level lying close to levels involved in the T$_3$--T$_6$ transitions. On the other hand, the unique valley-polarized nature of the $N{=}0$ level may play a role~\cite{CastroNeto:2009wq}. We note the values of \vf~at $\nu{=}\pm10$, where the $N{=}+2(-2)$ levels are filled (empty), largely agree with the $\nu{=}\pm6$ values, implying the decrease is in fact linked to filling the $N{=}0$ level.

The dashed lines in Fig.~\ref{vf} are calculated using a many-particle theory of graphene magneto-excitons~\cite{Bychkov:2008gi}. As shown, these predictions roughly capture the decrease of \vf~with increasing $|\nu|$, but are almost identical for T$_2$ and T$_3$ and do not reflect the observed plateaus about $\nu{=}0$ or the $|\nu|{=}4$ minima in T$_2$.

\begin{figure}[b]
\includegraphics[width=0.8\columnwidth]{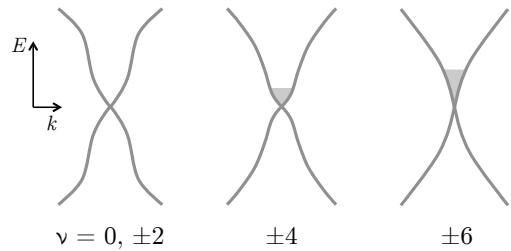}
\caption{``Zero-field'' graphene dispersion with slope---greatly exaggerated---given by \vf~values measured at $B{=}8$ T. The dispersions evolve with increasing filling factor (or density, indicated by shading) (see Supplemental Material~\cite{Note1}). \label{shift}} 
\end{figure}

\noindent (3) The several transitions recorded probe the graphene dispersion at energies up to $\pm300$ meV away from the Fermi level. In detail the dependence of \vf~on transition number evolves with $\nu$, and we have re-plotted these data to make this plain in the Supplemental Material~\cite{Note1}. Indeed, at $\nu{=}0$ the peak value of \vf~occurs for T$_3$, but as $\nu$ increases the peak shifts to lower transition numbers. We represent this schematically in Fig.~\ref{shift}, sketching a ``zero-field'' Dirac cone with slope \vf. This procedure results in an intriguing departure from the familiar linear dispersion. In contrast, the renormalized dispersion found by scanning tunneling spectroscopy remains linear over a comparable energy range about the Fermi level, with higher \vf~for lower densities~\cite{Chae:2012jk}. These divergent results can perhaps be reconciled by noting that in Ref.~\cite{Chae:2012jk}, electrons tunneled into a single LL of an interacting system, while here we are sensitive to the physics of an electron-hole pair with each particle in a separate LL having different index (and energy), noting that many-particle interactions differ for an exciton vs.~single electrons~\cite{Kallin:1985ud}.

\noindent (4) The \vf~found at $B{=}5$ and 11 T are qualitatively similar to those of Fig.~\ref{vf}, except the overall magnitude decreases with increasing $B$. This running of the velocity enhancement was predicted~\cite{Shizuya:2010go} and also recently observed in graphene magneto-Raman experiments~\cite{Faugeras:2015it}.

Returning to Fig.~\ref{trans}, T$_1$ exhibits a ${\approx}2.5$ meV splitting near $\nu{=}\pm2$. We speculate the $N{=}0$ level has become gapped due to coupling of the graphene and hbn lattices, which breaks the sublattice symmetry and generates a Dirac mass~\cite{Hunt:2013ef,Song:2013ji}; the splitting is due to transitions to both sides of the gapped level. Such lattice couplings also generate additional Dirac peaks on both sides of the main peak, which are not apparent in Fig.~\ref{transport}; however, these peaks shift to higher density for increasing graphene-hbn rotational misalignment, and thus may lie outside the $V_g$ range of this device. More work on samples with a range of relative graphene-hbn rotations is needed to verify this, but if correct, our method can be used to accurately determine the gap size and its dependence on rotation or other symmetry-breaking mechanisms.

The good quality of this sample leads to very narrow resonances with halfwidths falling below 3 meV for the sharpest T$_1$ transitions, leading to high-precision measurements that are likely to enable future spectroscopy of the fractal Landau levels underlying Hofstadter's butterfly~\cite{Hunt:2013ef,Dean:2013bv,Ponomarenko:2013hl}, broken symmetry states~~\cite{young:2012}, or even the fractional quantum Hall effect at lower temperatures~\cite{Dean:2011ks}. 

In conclusion, we have performed infrared magneto-spectroscopy on a high quality encapsulated monolayer graphene device. Several interband transitions provide direct evidence of contributions from many-particle physics, via a renormalized Fermi velocity that depends on the Landau level filling factor and the magnetic field strength. This work demonstrates that infrared magneto-spectroscopy can provide a novel tool for the study of correlated electron physics in graphene.

\begin{acknowledgments}
We thank D.~Abergel, J.~Balgley, C.~Dean, B.~Hunt, J.~Pollanen, and Li Yang for helpful discussions and experimental assistance. We acknowledge support from the Institute of Materials Science and Engineering at Washington University in St.~Louis.
\end{acknowledgments}

\bibliography{\refs}

\end{document}